\documentstyle[aps,prb,preprint]{revtex}
\begin{document}
\draft

\title{Band structure and optical anisotropy in V-shaped 
and T-shaped semiconductor quantum wires}

\author{G. Goldoni$^{1,2}$, F. Rossi$^{1,2}$, E. Molinari$^{1,2}$, and 
A. Fasolino$^{1,2,3}$}

\address{$^1$ Istituto Nazionale per la Fisica della Materia (INFM)}
\address{$^2$ Dipartimento di Fisica, Universit\`a di Modena, Via Campi
213/A, I-41100 Modena, Italy}
\address{$^3$ Institute of Theoretical Physics, University of Nijmegen,
Toernooiveld, 6525 ED Nijmegen, The Netherlands}

\date{\today}
\maketitle

\begin{abstract}

We present a theoretical investigation of the electronic and optical
properties of V- and T-shaped quantum wires. Valence
band mixing as well as realistic sample geometries are fully included 
through an accurate and efficient approach that is described here in detail.
We investigate the resulting valence band structure, which shows some 
significant peculiarities, such as an anomalously large spin splitting 
in the lowest heavy hole subband of T-shaped wires.
For both classes of wires we
obtain good agreement between calculated optical absorption and recent
experimental spectra, and we demonstrate that the analysis of optical
anisotropy can be used as an effective tool to extract information on
valence states, usually very difficult to obtain otherwise. 

\end{abstract}


\section*{introduction}

In recent years, one-dimensional (1D) semiconductor nanostructures have
received increasing attention. The potential technological application
of quantum wires (QWRs), e.g. in laser devices, has fueled a search for
new fabrication techniques and improved sample quality.~\cite{review}
In this area, recent investigations have focused on two classes of
structures, the so-called
V-shaped~\cite{Kapon89,Gailhanou93,Tiwari94,Rinaldi94a,Grundmann95} and
T-shaped~\cite{Pfeiffer90,Gershoni90,Stormer91,Goni92,Wegscheider95,Akiyama96a,Akiyama96b,Someya96}
QWRs (V-QWRs and T-QWRs). Due to high control on growth conditions and
strong confinement of the electron and hole wavefunctions on the scale
of few nanometers, V-QWRs and T-QWRs share desirable optical properties
for device applications, such as large exciton binding energy and small
linewidth. 

V-QWRs are obtained from a GaAs substrate grown along the [001]
crystallographic direction, patterned with [$\bar{1}$10]-oriented
V-shaped grooves obtained by chemical etching. The active region
consists of a GaAs layer cladded between two Al$_x$Ga$_{1-x}$As\ 
regions~\cite{Kapon89} or GaAs/AlAs superlattices~\cite{Rinaldi94a}
(SLs) overgrown on the patterned substrate. The confining potential
(see Fig.~\ref{fig:geom}) has a crescent shape profile. T-QWRs are
obtained by first growing a GaAs/Al$_x$Ga$_{1-x}$As\  SL on a (001) substrate.
After cleavage, a GaAs quantum well (QW) is grown over the exposed
(110) surface, resulting in a T-shaped active
region.~\cite{Pfeiffer90} In both cases, the electron and hole
wavefunctions are confined in the [001] and [110] crystallographic
directions, while the QWR free axis is parallel to the [$\bar{1}$10]
direction.

The optical spectroscopy of QWRs is more complex than for QWs of
similar lateral dimension, since in QWRs linewidths can be comparable
to intersubband splittings. On the other hand, a remarkable
peculiarity of QWRs with respect to QWs is that the optical activity
is strongly anisotropic when light is linearly polarized, with the
electric field directed parallel or perpendicular to the wire axis.
This has long been recognized to be a band structure effect due to the
quasi-1D character of electronic state, combined with heavy- and
light-hole (HH and LH) mixing.~\cite{Sercel90} The anisotropic
absorption is therefore used as a simple tool to reveal the 1D
character of electronic states in nanostructured materials.

In principle, the optical anisotropy can be exploited to single out
detailed information on the electronic states, since, as we will
show, it is very sensitive to specific details of the band structure.
In practice, this approach has been sofar limited by the lack of
realistic calculations for complex geometries, as the present V-QWRs
and T-QWRs. Indeed, common theoretical methods, even within
semi-empirical schemes as the tight-binding or the envelope function
approach, require a large scale computational effort. In order to keep
calculations tractable, up to now the optical properties of QWRs have
been investigated theoretically only for rather idealized structures
~\cite{Citrin89,Bock92a,Citrin92,Ando93,Goldoni95,Yamaguchi95},
yielding results that cannot be directly compared with experimental
spectra. Calculations have been performed for realistic QWR 
geometries~\cite{Rossi96a}, but they have sofar neglected HH-LH mixing.

Recently, we have demonstrated, by a combined theoretical-experimental 
study of V-QWRs,~\cite{our-apl}
that accurate band structure calculations for realistic 
structures provide quantitative predictions 
of photoluminescence excitation (PLE) spectra, 
and that detailed information on the valence band states can be 
singled out of the PLE anisotropy, despite the dominant role of the 
light conduction electrons in the optical spectra.
Such calculations were based on a 
recently devised method which provides the band structure 
for QWRs of arbitrary geometry at a relatively small computational
cost. The accuracy and the short computer times make
such calculations 
a practical characterization tool in conjunction with experimental results, 
as well as a predictive tool for new devices.

In this paper, we present a theoretical investigation of the electronic 
and optical properties of V-QWRs and T-QWRs.
%%% here%%%
%, extending our previous 
% studies~\cite{Rossi96a} to include valence band mixing. 
We focus on the relationship between optical anisotropy and band
structure, and we show how the analysis of optical anisotropy permits a
detailed spectroscopy of valence states, even when the large linewidth
of the spectra does not allow an identification  of the
valence-to-conduction subband transitions. For T-QWRs we also predict a
huge spin-splitting of the lowest valence subband, originating from the
interaction between the lowest HH levels of the intersecting QWs. 
For both classes of
wires, calculations are performed by the numerical method introduced in
Ref.~\onlinecite{our-apl} ---here described in detail---, thereby
demonstrating its accuracy and flexibility. 

The main approximation that is still present in our approach 
is the neglect of excitonic effects. Indeed, recent calculations 
which fully include Coulomb interaction in realistic QWR
profiles~\cite{Rossi96a} (but do not include HH-LH mixing) demonstrate that 
electron-hole interaction, besides giving rise to bound
excitonic states below the band edge, also modifies the
excitonic continuum above the band edge.
However, our previous investigation in V-QWRs~\cite{our-apl}
has shown that quantitative agreement is obtained 
between the observed PLE anisotropy, which probe the excitonic continuum, 
and the anisotropy in absorption spectra calculated 
neglecting excitonic effects. We interpret
this result as an indication that the electron-hole Coulomb
interaction, by mixing isotropically the (optically anisotropic)
electron and hole states, does not change the average anisotropy as
obtained by single particle band structure calculations. 
This is of course compatible with the possibility that 
the relative intensity of the absorption peaks for a given polarization
may be strongly affected.~\cite{Bock92a} 
The above arguments suggest that the approximation of neglecting 
electron-hole Coulomb coupling is a reasonable one for 
our purpose of studying optical anisotropies. 
Of course, its accuracy for the present V- and T-QWRs must be 
established {\rm a posteriori} by comparison with experiments, 
as we will do later on in this paper.

The theoretical background and the numerical method used in our
calculations are outlined in Sec.~\ref{sec:metodo}. Sections
\ref{sec:V} and \ref{sec:T} report the results of our calculations,
focusing on the band structure and optical anisotropy, for V-QWRs and
T-QWRs, respectively. 

\section{Modelling the band structure 
in wires of arbitrary geometry}
\label{sec:metodo}

In this section we describe the theoretical framework of our 
calculations. As we are interested in optical 
transitions near the direct band gap 
of GaAs, and because QWR electronic states are extended over several 
nanometers, we work within the envelope function approximation.
For the wires of interest in this paper, we choose the following Cartesian 
reference frame (see also Fig.~\ref{fig:geom}): 
$x$ along the [110] direction, $y$ along the [001] 
direction, and $z$ along the [$\overline{1}$10] direction. Therefore,
for both classes of wires, the QWR section
extends in the $x-y$ plane, while its free axis is parallel to the 
$z$-direction. Due to translational invariance, along this direction 
it is possible to define a 1D wavevector $k_z$.
In the $x$ and $y$ directions we assume a supercell periodicity, i.e. we 
describe arrays of QWRs. The size of the supercell can be taken large 
enough to describe effectively isolated wires when needed. 

Electron and hole states will be described separately by different 
effective mass equations. For conduction electrons,
we assume a single band approximation, which 
implies a parabolic energy dispersion in the free 
direction; the wavefunctions of the electron subbands are
\begin{equation}
\Psi^e_n({\bf r}) = F^e_n({\bf r}) \left|s,\sigma\right.\rangle,
\end{equation}
where $\left|s,\sigma\right.\rangle$ is the atomic $s$ state
with spin $\sigma\in\{\uparrow,\downarrow\}$,
$n$ is the subband index, and $F^e_n({\bf r})$ is the $n$-th
solution of the envelope function equation. Since we deal with a
two-dimensional confinement potential $V(x,y)$, we can factorize
$F^e_n({\bf r}) = e^{ik_zz} \psi^e_n(x,y)$,
where $\psi^e_n$ is the $n$-th solution of
\begin{equation}
\hat{H}^e \psi^e_n(x,y) = E^e_{n,k_z}\psi^e_n(x,y),
\end{equation}
with the electron effective mass Hamiltonian 
\begin{equation}
\hat{H}^e = 
\left[\frac{\hbar^2}{2m_e}\left(\hat{k}_x^2+\hat{k}_y^2+k_z^2\right)+V(x,y)\right];
\end{equation}
here $m_e$ is the electron effective mass, and $\hat{k}_x=-i\partial/
\partial x$, $\hat{k}_y=-i\partial/\partial y$. Of course, in this 
one-band description, the electron subbands $E^e_{n,k_z}$ depend quadratically 
on $k_z$:
\begin{equation}
E^e_{n,k_z} = \epsilon^e_n+\frac{\hbar^2k_z^2}{2m_e},
\end{equation}
where the $\epsilon^e_n$'s are the confinement energies (i.e., the 
subband edges). The subbands $E_{n,k_z}^e$  are spin-degenerate,
and the $\psi^e_n$'s are $k_z$-independent.

Conduction band states do not directly contribute to optical 
anisotropy, being mainly of isotropic $s$-type character, 
and in this case the above one-band description is sufficient.
Conversely, it is essential to give accurate account of $p$-type 
valence states by 
a multi-band description, in order to investigate optical anisotropy. 
Due to mixing between HH and LH subbands, 
both eigenvalues and envelope functions depend non-trivially on the 
wavevector $k_z$ and the hole subband index $m$. 
Using the compact notation $\alpha = (k_z,m)$, we write 
the hole wavefunctions 
\begin{equation}
\Psi^h_\alpha({\bf r}) = \sum_{J_m} F^h_{J_m,\alpha}({\bf r}) 
\left|3/2,J_m\right.\rangle,
\end{equation}
where $\left|3/2,J_m\right.\rangle$ 
are the four atomic states with total angular momentum 
$J=3/2$ and projection $J_m = +3/2, +1/2, -1/2, -3/2$ of $J$ along a 
quantization axis. Again, each envelope function $ F^h_{J_m,\alpha}({\bf r}) $ 
can be factorized as 
$ F^h_{J_m,\alpha}({\bf r}) = e^{ik_zz}\psi^h_{J_m,\alpha}(x,y)$, 
and the four-component vector
$\mbox{\boldmath$\psi$\unboldmath}^h_\alpha (x,y) = (\psi^h_{+3/2,\alpha}, 
\psi^h_{-1/2,\alpha}, 
\psi^h_{+1/2,\alpha},
\psi^h_{-3/2,\alpha})$ is the $m$-th solution at point $k_z$ 
of the multi-band effective mass equation~\cite{nota-labels}
\begin{equation}
\hat{{\bf H}}_L \mbox{\boldmath$\psi$\unboldmath}^h_\alpha = 
E^h_\alpha \mbox{\boldmath$\psi$\unboldmath}^h_\alpha ,
\label{eq:multiband}
\end{equation}
where $\hat{{\bf H}}_L$ is the Luttinger Hamiltonian.~\cite{Luttinger56}
With the above choice of coordinate axes and the 
quantization axis of ${\bf J}$ along the [110] direction, 
following Ref.~\onlinecite{Xia91} the Luttinger Hamiltonian reads 
\begin{equation}
\hat{{\bf H}}_L = \frac{\hbar^2}{2m_0}\left(
\begin{array}{cccc}
\hat{P}_1 + V(x,y) & \hat{R} & \hat{Q} & 0 \\
\hat{R}^\dagger & \hat{P}_2 +V(x,y) & 0 & -\hat{Q} \\
\hat{Q}^\dagger & 0 & \hat{P}_2 +V(x,y)& \hat{R} \\
0 & -\hat{Q}^\dagger & \hat{R}^\dagger & \hat{P}_1 + V(x,y)
\end{array}\right)
\begin{array}{c}
\left|3/2,+3/2\right.\rangle \\
\left|3/2,-1/2\right.\rangle \\
\left|3/2,+1/2\right.\rangle \\
\left|3/2,-3/2\right.\rangle 
\end{array},
\label{eq:H_L}
\end{equation}
where
\begin{mathletters}
\begin{eqnarray} 
\hat{P}_1 & = & \left(\gamma_1-\frac{\gamma_2+3\gamma_3}{2}\right) \hat{k}_x^2 +
          \left(\gamma_1+\gamma_2\right) \hat{k}_y^2 + 
          \left(\gamma_1-\frac{\gamma_2-3\gamma_3}{2}\right) k_z^2  ,
          \\
\hat{P}_2 & = & \left(\gamma_1+\frac{\gamma_2+3\gamma_3}{2}\right) \hat{k}_x^2 +
          \left(\gamma_1-\gamma_2\right) \hat{k}_y^2 + 
          \left(\gamma_1+\frac{\gamma_2-3\gamma_3}{2}\right) k_z^2 ,
          \\
\hat{R} & = & -\sqrt{3}\left[
           -\frac{\gamma_2-\gamma_3}{2}\hat{k}_x^2 + 
           \left(\gamma_2 \hat{k}_y-2i\gamma_3 k_z\right) \hat{k}_y -
           \frac{\gamma_2+\gamma_3}{2}k_z^2
         \right] ,
         \\
\hat{Q} & = & -2\sqrt{3}\left(\gamma_3 \hat{k}_y-i\gamma_2 k_z\right) \hat{k}_x .
\end{eqnarray}
\end{mathletters}
This Hamiltonian provides the (positive definite) hole 
subbands, referred to the bulk valence band edge,
as a function of the in-wire wavevector $k_z$,  
including HH-LH mixing. The hole 
Hamiltonian for structures grown along crystallographic directions 
different from the present ones can be obtained along the lines of 
Ref.~\onlinecite{Xia91}. In the above electron and hole effective mass 
Hamiltonians 
we neglect the material dependence of the electron effective mass $m_e$
and the Luttinger parameters $\gamma_1$, 
$\gamma_2$, and $\gamma_3$, and we always use the bulk GaAs values listed in 
Table~\ref{tab:parameters}. 
Although our approach can be extended to account for 
the material dependence, this would be a small effect in our calculations.

One possible approach to solve the multi-band equation
(\ref{eq:multiband}) is to split the problem in two parts: the first
step is to solve the two Schr\"odinger-like equations arising from the
diagonal terms of (\ref{eq:H_L}),
\begin{eqnarray}
\left[\hat{P}_1+V(x,y)\right]\phi^1_i(x,y) = \epsilon^1_i\phi^1_i(x,y), \\
\left[\hat{P}_2+V(x,y)\right]\phi^2_i(x,y) = \epsilon^2_i\phi^2_i(x,y). 
\end{eqnarray}
The second step is to use the set $\phi^1_i$ as a basis to expand
the components $\pm3/2$ of the vector 
$\mbox{\boldmath$\psi$\unboldmath}^h_\alpha$, and the set 
$\phi^2_i$ as a basis for
the components $\pm1/2$. In this representation, the 
diagonal matrix elements of the ${\bf H}_L$ are given 
by the two sets of scalar numbers $\epsilon^1_i$, $\epsilon^2_i$, and 
matrix elements need only to be calculated for off-diagonal terms, 
using the functions $\phi^1_i$, $\phi^2_i$. This approach has the 
following drawback: since the effective masses appearing in 
$\hat{P}_1$, $\hat{P}_2$ (the so-called HH and LH effective masses) 
are very different, the two sets of eigenvalues $\epsilon^1_i$, $\epsilon^2_i$ 
span different energy ranges. The set $\epsilon^2_i$, 
being the spectrum of a light particle, will have larger gaps,
and the ground state will be higher in energy
than for the set $\epsilon^1_i$. Since we are mainly interested 
in the low-lying hole subbands near $\epsilon^1_1$,
this representation, although exact 
in principle, is poorly convergent with respect to the number of states 
$\phi^1_i$, $\phi^2_i$ included in the basis, and it is not practical in 
numerical calculations. 

To solve this problem, we propose a scheme in which 
we expand the components of $\mbox{\boldmath$\psi$\unboldmath}_\alpha$ using 
the solutions of two 
Schr\"odinger-like equations with two fictitious, arbitrary masses
$m^+$, $m^-$,
which then we tune in order to improve the convergence:
\begin{mathletters}
\begin{eqnarray}
\left[\frac{\hbar^2}{2m^+}\left(\hat{k}_x^2+\hat{k}_y^2\right)+V(x,y)\right]
\phi^+_\nu (x,y) & = & \epsilon^+_\nu\phi^+_\nu (x,y), 
\label{eq:fict-f}\\
\left[\frac{\hbar^2}{2m^-}\left(\hat{k}_x^2+\hat{k}_y^2\right)+V(x,y)\right]
\phi^-_\mu (x,y) & = & \epsilon^-_\mu\phi^-_\mu (x,y)
\label{eq:fict-l}
\end{eqnarray}
\end{mathletters}
(here and in the following we use the index $\nu$ for $+$ states, and $\mu$  
for $-$ states).
In this way, we diagonalize exactly only {\em part} of the kinetic energy terms 
$\hat{P}_1$ and $\hat{P}_2$ of $\hat{{\bf H}}_L$, but the potential $V(x,y)$ 
is exactly diagonalized. Of course, in this representation 
$\epsilon^+_\nu$, $\epsilon^-_\mu$ are {\em not} 
the diagonal elements of the matrix representing the 
Hamiltonian, and $\hat{P}_1$, $\hat{P}_2$ give rise to additional off-diagonal 
terms. However, the time spent to calculate the
additional terms is more than compensated by the improved 
convergence which can be achieved by properly choosing $m^+$ and $m^-$.
In the end, we shall find it convenient 
to choose $m^+=m^-$, and equal to the heavy hole 
effective mass along the [001] 
direction. 

To implement this idea, we add and subtract a term $\hbar^2/2m^+(\hat{k}_x^2+
\hat{k}_y^2)$ to $\hat{P}_1$, and a term $\hbar^2/2m^-(\hat{k}_x^2+\hat{k}_y^2)$ 
to $\hat{P}_2$. Then we obtain:
\begin{mathletters}
\begin{eqnarray}
\hat{P}^+ & = & \hat{P}_1 = \frac{1}{m^+}\left(\hat{k}_x^2+\hat{k}_y^2\right)+
          p^+_x \hat{k}_x^2+p^+_y \hat{k}_y^2 + 
          \left(\gamma_1-\frac{\gamma_2-3\gamma_3}{2}\right) k_z^2  ,
          \\
\hat{P}^- & = & \hat{P}_2 = \frac{1}{m^-}\left(\hat{k}_x^2+\hat{k}_y^2\right)+
          p^-_x \hat{k}_x^2+p^-_y \hat{k}_y^2 + 
          \left(\gamma_1+\frac{\gamma_2-3\gamma_3}{2}\right) k_z^2  ,
\end{eqnarray}
\end{mathletters}
where
\begin{mathletters}
\begin{eqnarray}
p^+_x & = & \left(\gamma_1-\frac{\gamma_2+3\gamma_3}{2}\right)-\frac{1}{m^+}, \\
p^+_y & = & \left(\gamma_1+\gamma_2\right)-\frac{1}{m^+}, \\
p^-_x & = & \left(\gamma_1+\frac{\gamma_2+3\gamma_3}{2}\right)-\frac{1}{m^-}, \\
p^-_y & = & \left(\gamma_1-\gamma_2\right)-\frac{1}{m^-}. 
\end{eqnarray}
\end{mathletters}

Then we solve the equations (\ref{eq:fict-f})-(\ref{eq:fict-l})
by a plane-wave 
expansion, as outlined in Ref.~\onlinecite{Rossi96a}. Typically,
we fix two energy cutoffs, $E^+_{cut}$ and 
$E^-_{cut}$, and we find the $N^+$ and $N^-$ eigenstates 
which fall below the cutoffs. 
Using the eigenfunctions $\phi^+_\nu$, $\phi^-_\mu$, 
we form the following basis set:
\begin{mathletters}
\begin{eqnarray}
\left|+,\nu,\uparrow\rangle\right. = \left(\begin{array}{c} 
                    \phi^+_\nu \\ 0 \\ 0 \\ 0 
                  \end{array}\right),
\left|+,\nu,\downarrow\rangle\right. = \left(\begin{array}{c} 
                      0 \\ 0 \\ 0 \\ \phi^+_\nu 
                    \end{array}\right), 
\label{eq:basis-f}\\
\left|-,\mu,\uparrow\rangle\right. = \left(\begin{array}{c} 
                    0 \\ \phi^-_\mu \\ 0 \\ 0
                  \end{array}\right),
\left|-,\mu,\downarrow\rangle\right. = \left(\begin{array}{c} 
                      0 \\ 0 \\ \phi^-_\mu \\ 0 
                    \end{array}\right),
\label{eq:basis-l}
\end{eqnarray}
\end{mathletters}
with $\nu=1\dots N^+$ 
and $\mu=1\dots N^-$, 
and we expand $\mbox{\boldmath$\psi$\unboldmath}^h_\alpha$ in this basis:
\begin{equation}
\mbox{\boldmath$\psi$\unboldmath}^h_\alpha = 
  \sum_{\nu\sigma} C^+_\alpha(\nu,\sigma) \left|+,\nu,\sigma\right.\rangle +
  \sum_{\mu\sigma} C^-_\alpha(\mu,\sigma) \left|-,\mu,\sigma\right.\rangle.
\end{equation}
The explicit matrix elements of $\hat{{\bf H}}_L$ in this basis are given in 
appendix \ref{sec:app-a}. 
The total dimension of the Hamiltonian matrix in this representation 
is $2\times(N^++N^-)$. 
All we need to compute, in order to evaluate the 
matrix elements, are integrals of the kind
\begin{mathletters}
\begin{eqnarray}
\int_\Omega \left[\phi^{\gamma}_i(x,y)\right]^*  & \hat{k}_\beta &
\phi^{\gamma^\prime}_j(x,y)
\,dx\,dy,
\\
\int_\Omega \left[\phi^{\gamma}_i(x,y)\right]^*  & \hat{k}_\beta & \hat{k}_{\beta^\prime} 
\phi^{\gamma^\prime}_j(x,y) 
\,dx\,dy,
\end{eqnarray}
\end{mathletters}
evaluated over the supercell volume $\Omega$,
where $\beta, \beta^\prime \in \{x,y\}$ and $\gamma, \gamma^\prime\in\{+,-\}$;
these are easily obtained given the  plane-wave expansions of the
$\phi^+_i$'s and $\phi^-_i$'s.
The choice of $m^+$, $m^-$, which is important in order to obtain an
efficient convergence, is discussed in detail in appendix
\ref{sec:app-b}; here we  only anticipate that in all our calculations we
use $m^+=m^-=(\gamma_1-2\gamma_2)^{-1}$. 

Once we have calculated the electron and hole subbands by the above 
method, we are in the position to evaluate the absorption spectrum 
$\alpha_\epsilon(\hbar\omega)$ in 
the dipole approximation, summing 
the dipole matrix element, with the appropriate polarization of 
light $\epsilon$, over all electron and hole states:
\begin{equation}
\alpha_\epsilon(\hbar\omega) \propto \sum_{\alpha,n,\sigma} 
  \left|M^\epsilon_{\alpha\rightarrow n,\sigma}\right|^2 
  \delta(E^e_n+E^h_\alpha+E_g-\hbar\omega),
\label{eq:abs}
\end{equation}
where the optical matrix elements $M^\epsilon_{\alpha\rightarrow n,\sigma}$
are given in appendix \ref{sec:app-c},
and $E_g$ is the bulk energy gap of GaAs; tipically, a set of sixty
$k_z$ points have been included in the summation.
The absorption spectra shown in this paper have been obtained by 
superimposing 
a gaussian broadening $\sigma_b$ to $\alpha_\epsilon(\hbar\omega)$,
in order to simulate the inhomogenous 
broadening due to structural imperfections of the samples. The 
broadened spectrum is obtained as 
\begin{equation}
\alpha_\epsilon(\hbar\omega) = \int^\infty_{-\infty}
\alpha_\epsilon(\hbar\omega^\prime)
  e^{-\hbar^2(\omega-\omega^\prime)^2/2\sigma_b^2} d\omega^\prime.
\end{equation}

\section{Band structure and optical properties of V-shaped wires}
\label{sec:V}

\subsection{Samples}

As a prototype of V-QWRs, we first consider a sample described in
Refs.~\onlinecite{Rinaldi94a,Rinaldi94b}. This consists of an active GaAs
layer embedded in a (AlAs)$_4$/(GaAs)$_8$ SL, overgrown by molecular
beam epitaxy on the exposed surface of the etched substrate. As in
Refs.~\onlinecite{our-apl,Rinaldi94b}, we use the V-shaped potential
profile obtained by digitalizing a TEM micrograph of the sample. We
also adopt the same supercell geometry as in
Refs.~\onlinecite{our-apl,Rinaldi94b}. The complicated structure of
the SL which provides the quasi-1D confinement is modelled by a
homogeneous barrier with effective conduction and valence band
offsets, $V^e_{\mbox{\tiny eff}}$ and $V^h_{\mbox{\tiny eff}}$,
respectively. Based on previous
investigations~\cite{Rinaldi94b,our-apl} for the same sample, we take
$V^e_{\mbox{\tiny eff}}=150\,\mbox{meV}$ and $V^h_{\mbox{\tiny
eff}}=85\,\mbox{meV}$. It should be noted that the effort of including
exactly the confining SL in the calculations would not necessarily
result in improved accuracy, as the envelope function approximation
itself looses its validity for such short-period SLs.

In order to investigate the role of the confinement in the optical
properties, we shall consider two sample profiles, which differ in the value of 
the confinement length $L$ at the bottom of the V-shaped region 
(see Fig.~\ref{fig:geom}): 
profile A ($L = 8.7\,\mbox{nm}$) and profile B ($L = 6.83\,\mbox{nm}$). 
These SL-embedded QWRs will be labelled A/SL and B/SL.

A key issue which make nm-scale QWRs interesting for electro-optical
applications are large confinement energies which can be obtained with
large band-offsets in addition to geometric confinement. In view of
this fact, we shall compare the samples described above with samples
having the same profiles and barriers constituted by pure AlAs, that we
will label A/AlAs and B/AlAs. The parameters of the four samples are
summarized in Table~\ref{tab:samples}. 

\subsection{Band structure}

A qualitative interpretation of the band structure of a V-QWR 
can be obtained by adding to a QW of width $L$ an additional lateral
confinement due to the crescent shape of the profile. In 
the lowest approximation, the latter can be thought of as a parabolic 
potential~\cite{review} which splits each subband of the parent QW into a 
new set of subbands.
Since the additional confinement is less effective 
than the confinement due to the original QW, the new
sets of subbands have smaller gaps with respect to the subband splittings 
of the parent QW. This simplified picture will serve as a guideline 
for the discussion of numerical results obtained for the 
actual samples. 

The calculated energies at $k_z=0$ of the lowest 
conduction and valence states are listed for reference in 
Table \ref{tab:V-edges} for the four samples.
In the following we shall focus on hole 
subbands, which are shown in Fig.~\ref{fig:V-bande} (right panel)
for sample A/SL. 
First, we note that at $k_z=0$ each subband is doubly degenerate, while 
at finite $k_z$
the subbands are spin-split, due to the lack of inversion symmetry of 
the confining potential.~\cite{nota-spin} 
Splittings are in the range of few meV. 
Secondly, a strongly non-parabolic energy 
dispersion is evident. This fact, familiar from QWs, is due to mixing 
of states with HH ($J_m=\pm3/2$) character and LH ($J_m=\pm1/2$) character. 

The HH/LH character of hole states influences the optical 
properties of the sample, since different atomic orbital components have 
different oscillator strenghts; of particular interest from this point
of view, are the $k_z=0$ states which, due to the large density of states (DOS) 
stemming from their quasi-1D character, mainly contribute to the 
absorption intensity. 
Contrary to the case of QWs, hole subbands cannot be 
strictly classified as HH and LH states even at $k_z=0$.
To define the HH/LH character in the 
present samples, we note that the direction of strongest 
confinement is the [001] direction, as demonstrated by charge density 
maps of the lowest lying states which we will show later in 
Fig.~\ref{fig:V-chd}. It is therefore  
meaningful to calculate the HH/LH character along this direction, 
because this would be the quantization axis 
of $J$ in an equivalent [001]-grown QW of width $L$. 
To do this, we calculate the rotated vector 
$\mbox{\boldmath$\psi$\unboldmath}^h_{\alpha,R}(x,y)  = {\bf R}^{-1}\cdot\mbox{\boldmath$\psi$\unboldmath}^h_{\alpha}(x,y)$, 
where 
\begin{equation}
{\bf R} = \frac{1}{2\sqrt{2}}
        \left(\begin{array}{cccc}
          1 & -\sqrt{3} & \sqrt{3} & 1 \\
          \sqrt{3} & 1 & -1 & \sqrt{3} \\
          \sqrt{3} & 1 & 1 & -\sqrt{3} \\
          1 & -\sqrt{3} & -\sqrt{3} & -1 
        \end{array}\right).
\end{equation}
${\bf R}$ is obtained by diagonalizing the matrix ${\bf J}_y$ written in the 
representation in which ${\bf J}_x$ is diagonal, with eigenvalues $J_m$.
Then we define the HH- and LH-projected charge densities
\begin{mathletters}
\begin{eqnarray}
\label{eq:rho-HH}
\rho_\alpha^{HH} (x,y)&=&\sum_{J_m=\pm3/2}\left|\psi^h_{\alpha,R,J_m}(x,y)\right|^2, \\
\label{eq:rho-LH}
\rho_\alpha^{LH} (x,y)&=&\sum_{J_m=\pm1/2}\left|\psi^h_{\alpha,R,J_m}(x,y)\right|^2.
\end{eqnarray}
\end{mathletters}
Finally, the HH and LH character is obtained by integrating the above charge 
densities over all space. (In the above equations, the 
real-space representation is chosen for clarity. The corresponding 
expressions in Fourier space, which we use in the numerical 
implementation, 
are very easy to obtain and are not explicitated here.)

In the left panel of Fig.~\ref{fig:V-bande} we show the calculated 
LH character of the hole subbands at $
k_z=0$. This is best analized in connection with Fig.~\ref{fig:V-chd}, 
where we show the total and projected
charge densities of the hole states at $k_z=0$ for the same A/SL sample.
In the ground state, the LH component is rather 
small ($\sim8\%$), but it increases rapidly for the excited subbands. 
Correspondingly, the ground state is well localized
[Fig.~\ref{fig:V-chd}(a)], 
while the lowest 
excited states [Figs.~\ref{fig:V-chd}(b), \ref{fig:V-chd}(c)]
have wavefunctions which increasingly extend along the 
V-QWR sidewalls. The regular 
increase of the LH component is interrupted by the $m=25,26$ levels 
(13th doublet: 
recall that each point at $k_z=0$ is doubly degenerate) at $\sim27\
\mbox{meV}$ which is mainly of LH character ($56\%$). 
This can be interpreted as the reminescent of 
the LH state of a QW with confinement length $L$. The
wavefunction of this level [Fig.~\ref{fig:V-chd}(d)], 
in fact, is again well localized, analogously to the 
ground state, and in contrast to the wavefunctions of nearby states.
We shall comment later on the fingerprints of this strongly LH-like state in 
the optical spectra, and, particularly, in the optical anisotropy. Note 
also that in Fig.~\ref{fig:V-bande} 
there are other ``jumps'' in the LH character at higher energies; 
these correspond to energies where ladders of levels of different 
simmetries, like an additional nodal plane parallel to the [110] 
direction, begin.

In Fig.~\ref{fig:sl-vs-bk} we compare the LH character vs subband
energy at $k_z=0$ for samples with different barriers (SL and pure
AlAs). For both profiles A and B, the strong confinement due to AlAs
barriers induces not only the expected blue-shift of the subbands, but
also reduces the LH character of the lowest LH-like state, as compared
to SL barriers: in sample A/SL, the $m=25,26$ doublet is 56\%
LH, while its counterpart in sample A/AlAs, the $m=27,28$ doublet, is
only 42\% LH. In sample B/SL, the $m=35,36$ doublet is 75\%
 LH, while its counterpart in sample B/AlAs, the $m=37,38$
doublet, is only 52\% LH.

\subsection{Optical properties}

In Fig.~\ref{fig:V-abs} we show the calculated absorption intensity for the 
four samples of Table~\ref{tab:samples} and for light linearly polarized 
parallel to the wire axis, $I_\parallel$, 
and perpendicular to it along the [110] direction, $I_\perp$.
A gaussian broadening of $\sigma_b=\pm4.5\,\mbox{meV}$ 
has been included.~\cite{our-apl} For all samples, we also report 
the relative optical anisotropy (thick 
lines in Fig.~\ref{fig:V-abs}), defined 
as $100*(I_\parallel-I_\perp)/I_\parallel$. The 
optical absorption spectra are obtained integrating over the whole band 
structure (i.e., integrating over $k_z$ and summing over electron and
hole subbands and spin);
however, for illustration we show 
the strongest optical transitions due to $k_z=0$ states,
and for one spin orientation of the conduction electrons, 
for light polarization perpendicular to the QWR axis
(histograms in  Fig.~\ref{fig:V-abs}).
To identify the electron and hole states involved in each transition, 
we use the label $n/m$, where $n$ and $m$ are, 
respectively, the indices of the conduction 
and valence states.

We first focus on sample A/SL, for which experimental data are 
available.~\cite{our-apl}
The calculated anisotropy in the low energy part ($\lesssim 1.62\,\mbox{eV}$) 
is $10\div20\%$, with a deep minimum at $\sim1.59\,\mbox{eV}$, where the 
anisotropy is almost suppressed. Both the average anisotropy and the 
position of the minimum are in {\em quantitative } agreement with experimental 
data.~\cite{our-apl} The agreement worsens at higher energies 
($\gtrsim 1.62\,\mbox{eV}$), 
where the calculated anisotropy drops rapidly and
finally changes sign, while experimental data~\cite{our-apl} show an increase. 
We believe that this discrepancy is due to our ``effective''
description of the barriers which affects particularly the higher-lying
hole states.

By comparing the calculated anisotropy of the four samples, it appears 
that a more or less pronounced dip over a range of $\sim10\,\mbox{meV}$
(i.e., the linewidth of the broadened spectra) 
is always present in the low energy range of the spectrum, 
superimposed onto a background of an otherwise large and 
positive anisotropy. Additional structure, particularly for sample 
A/AlAs, is also present in the high energy range. Note also that the 
maximum anisotropy is in the range $15\div25\%$, and does not change 
dramatically in the different samples.

By studying the optical matrix elements, it can be shown that the dips
in the anisotropy are due to states with a large LH
character,~\cite{Bock92a} e.g.,
the $m=25$ subband in sample A/SL. Since this is a localized state and
has a large spatial overlap with the first conduction subband, it
contributes to the low energy part of the spectrum. This is true in
general: as can be seen from Fig.~\ref{fig:V-abs}, for all samples the
anisotropy dip corresponds to a $k_z=0$ transition (highlighted as
black bars in the histograms) between the electron ground state
($n=1$) and an excited hole state which is the $m=25$, the $m=27$, the
$m=36$, and the $m=37$ level for A/SL, A/AlAs, B/SL, and B/AlAs,
respectively. For sample A/SL, comparison with Fig.~\ref{fig:V-bande}
shows that the involved $m=25$ hole state has a strong LH character
and its wavefunction (Fig.~\ref{fig:V-chd}(d)) is strongly localized.
The large LH component makes that the intensity for the two
polarizations is reversed with respect to the strongly HH-like ground
state, causing the dip in the anisotropy at $1.59\,\mbox{meV}$. A
similar correspondence between the dip and a localized LH-like state
applies also to the other samples.

An immediate consequence of the above result is that, since both the
ground HH state and the LH-like state couple with the lowest electron
subband, {\em the difference in energy between the onset of the
continuum and the position of the dip in the anisotropy is a direct
measure of the energy splitting between the ground HH and the first LH
state, independently of the electron confinement}. Note that such
information cannot be extracted from the absorption spectra alone, as
the large broadening prevents the identification of any single
transition apart from the fundamental one.

The HH/LH splitting obtained in the above manner can be used to
extract band structure parameters as, for example, the effective hole
confinement $V^h_{\mbox{\tiny eff}}$. In the measured PLE spectra for
sample A/SL,~\onlinecite{our-apl} the anisotropy dip lies $\sim
16\,\mbox{meV}$ above the onset of the continuum; in
Fig.~\ref{fig:e-hh-lh} we report the HH/LH splitting calculated for
several values of $V^h_{\mbox{\tiny eff}}$ (full dots) for this
sample. We also show, for comparison, the HH/LH splitting calculated
by a simple square well model for a QW of width $L$ (empty dots),
using the bulk HH and LH effective mass along [001] . It can be seen
that the splitting is quite sensitive to $V^h_{\mbox{\tiny eff}}$, and
that the experimental value is compatible with $V^h_{\mbox{\tiny
eff}}\sim 80\,\mbox{meV}$. This observation provides a good criterion
for choosing the confinement energy of holes, which would otherwise be
rather arbitrary. Using this procedure, $V^h_{\mbox{\tiny eff}}$ was
finally taken equal to $85\,\mbox{meV}$ in Ref.~\onlinecite{our-apl}.
Note that a variation of the valence band offset of
$\pm15\,\mbox{meV}$ shifts the HH/LH splitting of $\pm
1.3\,\mbox{meV}$, well within the experimental accuracy for the
determination of the HH/LH splitting.~\cite{our-apl} Note also that,
for a rough estimation of $V^h_{\mbox{\tiny eff}}$, calculations using
a simple square well model may be sufficient.

The comparison with the absorption spectra demonstrates that $k_z=0$
transitions alone give a poor estimate of the integrated spectrum.
This is due to the DOS contribution of the hole subbands which are
strongly non parabolic, and have a large DOS also for $k_z$-points
away from $k_z=0$ (see Fig~\ref{fig:V-bande}). Finally, note the
relaxation of selection rules for $k_z=0$ transitions shown in
Fig.~\ref{fig:V-abs}. Indeed, if the envelope-function
parity-conserving selection rule would be obeyed, only transitions of
the type $n/(m=2n)$ or $n/(m=2n-1)$ (depending on electron spin orientation)
would be allowed.

\section{Band structure and optical properties of T-shaped wires}
\label{sec:T}

\subsection{Samples}

We investigate a T-QWR with the geometry sketched in Fig.~\ref{fig:geom}.
In our calculation, 
the SL grown along the [001] direction is constituted by 5.3 nm 
wide QWs (QW1), separated from each other by 50 nm wide AlAs barriers, 
while the QW grown along the [110] direction (QW2) is 4.8 nm 
wide. These parameters correspond to a sample for which 
polarization-dependent PLE spectra are available.~\cite{Akiyama96a}
 
Note that the T-QWRs which form at the intersection between QW1 
and QW2 are not uncoupled in the [001] 
direction, due to the SL structure of QW1s, a fact which we fully take 
into account in the supercell representation used in our 
calculations. On the other hand, the T-QWRs are isolated along the [110] 
direction; in our calculations, this is simulated by truncating QW1
with a AlAs barrier 50 nm on the left hand side. 

\subsection{Band structure}

The nature of the 1D confinement in T-QWRs is rather different with
respect to the V-QWRs case. There, we used the picture of a lateral
confinement added on a QW structure which, therefore, localizes all
states  in the $x-y$ plane. Conversely, T-QWR states can be
better interpreted as the result of the coupling between 2D states of
the QWs, QW1 and QW2. Beside perturbing the 2D states, this coupling
also induces localized states or resonances; these can be also seen as
due to a 1D defect in the otherwise translationally invariant (in the
QW plane) 2D states of a QW, due to the coupling with the other QW.
This picture has non trivial consequences, particularly for hole
states, which we analyze in the following.

First, we focus on the lowest lying states at $k_z=0$. Electrons and holes are
very different from this point of view: contrary to the electron
effective mass which is isotropic in GaAs, the HH effective mass is
strongly anisotropic between the [001] direction ($m_{HH}=0.377$) and
the [110] direction ($m_{HH}=0.69$). The effect of this difference is
shown in Fig.~\ref{fig:T-chd-1}, where we report the total charge densities
for the lowest electron and hole states at $k_z=0$. For the present
structure, being QW1 and QW2 of comparable width, the electron ground
state [Fig.~\ref{fig:T-chd-1}(a)] is a quasi-1D state extending both in
QW1 and QW2; the lowest excited subbands
[Figs.~\ref{fig:T-chd-1}(b),(c),(d)]
belong to the continuum of 2D states of QW1, since QW1 is wider
(actually, discrete levels are obtained here, due to the supercell
method); the localized quasi-1D electron lies 15 meV below the QW1
continuum. 

On the contrary, for holes the larger mass along [110] more than
compensate for the smaller width of QW2, and the hole ground state
[Fig.~\ref{fig:T-chd-1}(a)], although localized in the center of the T,
is much more a QW2-like state, only weakly perturbed by coupling to
QW1; accordingly, the lowest excited hole subbands
[Fig.~\ref{fig:T-chd-1}(b),(c)] are basically 2D states belonging to
QW2, up to the fourth level [Fig.~\ref{fig:T-chd-1}(d)], which is a
QW1 state. The quasi-1D ground state is practically degenerate with
the QW2 continuum; note however, that, in contrast to conduction
electrons, there is a second hole level [Fig.~\ref{fig:T-chd-1}(b)]
with a significant component along QW1.

%In the present single particle
%picture, therefore, these excited states have small overlap and,
%consequently, small oscillator strength for absorption of light. We
%can also argue that Coulomb interaction will tend to oppose to this
%different localization and, therefore, the neglect of excitonic
%effects is expected to be a more serious limitation in T-QWR than in
%V-QWR.

The full band structure of holes is shown in the right-hand panel of 
Fig.~\ref{fig:T-bande}. In the rather complex dispersion of the 
subbands, we can distinguish a peculiar feature of T-QWR, namely a huge 
spin-splitting of the lowest doublet at finite $k_z$ which, 
at $k_z\simeq0.035\,\mbox{\AA$^{-1}$}$, is $\sim15\,\mbox{meV}$ 
for the present structure. 

Before discussing the origin of  such a large spin-splitting, it is
useful to examine the actual localization of the calculated 
states in the QWR. To this aim, we need to discriminate the states peaked
around the center of the T from states that are typical of one of 
the parent QWs and are 
left essentially unchanged by the interaction with the other one.
This is particularly useful because we expect that most of the states 
of Fig.~\ref{fig:T-bande} simply arise from folding of QW states induced by 
the supercell periodicity and to the truncation of the T along the [110] direction. 
Therefore, in  Fig.~\ref{fig:T-bande} we 
identify by full circles those states which, by direct inspection of
the wavefunction, show a strong localization in the center of the T.
Indeed, it appears that only a limited subset 
has predominantly a localized character, and can 
therefore be assigned to quasi-1D QWR-like states or resonances. 
These include the localized states arising from the lowest two 
doublets, already discussed above, 
as well as the resonant states falling around 40 meV. 
The remaining states are similar to 2D QW-like states,
with charge density localized mostly in QW1 or QW2.

To clarify the origin of the large spin-splitting, it is useful to
compare the band structure of the T-QWR with the hole subbands of
the parent isolated QW1 and QW2. In Fig.~\ref{fig:T-bande} we show with
open squares and open circles the lowest HH subband of QW1 and QW2; 
the LH subbands, for these thicknesses, lie high in energy
($>80\,\mbox{meV}$ in QW1 and $>100\,\mbox{meV}$ in QW2).~\cite{Los95} 
A small gap of $\sim 8\, \mbox{meV}$ separates the two levels at $k_z=0$; 
as already noted above, the lowest state is QW2, due to the larger HH mass. 
Owing to different HH-LH mixing in each QW separately, 
the HH subbands of the two parent QWs have different energy 
dispersion as a function of $k_z$ and cross at some finite $k_z$.

Large spin splittings are in general an effect of HH-LH mixing in 
asymmetric structures.~\cite{Bastard} 
In the present case, however, the energy difference between 
the HH and LH levels in each isolated QW is too large to explain the huge
splitting that we have found in the lowest energy subbands of the QWR.
Naively, one would rather expect small gaps (induced only indirectly 
by coupling to the far-lying LH states) to open at the crossing of the QW1 
and QW2 subbands.
Note, however, that the HH states of the parent 
QWs are eigenstates of ${\bf J}$ with $J_m=\pm3/2$, 
but {\em with the quantization axis along different directions}. 
Taking, for example, the ${\bf J}$ quantization axis along [110], a HH 
state of QW2 is $\left|QW2\right.\rangle=
\left|3/2,+3/2\right.\rangle$ (for one of the 
degenerate spin orientations). A HH state of QW1 {\em written in the 
same basis}, is 
\begin{equation}
\left|QW1\right.\rangle=\frac{1}{2\sqrt{2}}
\left(\left|3/2,+3/2\right.\rangle+\sqrt{3}\left|3/2,+1/2\right.\rangle+
      \sqrt{3}\left|3/2,-1/2\right.\rangle+\left|3/2,-3/2\right.\rangle\right).
\end{equation}
In words, a HH state of QW1 has a strong LH component from the point
of view of QW2. Therefore, the lowest HH subbands arising from QW1 and
QW2 are strongly coupled, which results in a strong avoided-crossing
behaviour and a very large spin-splitting.

This interpretation is supported by our calculation of the LH
character, which is shown in the left panel of Fig.~\ref{fig:T-bande}.
In view of the fact that the hole ground state is rather QW2-like,
we have computed the LH character 
with the quantization axis along [110], i.e., it is
obtained, as in Sec.~\ref{sec:V}, by integrating the projected
charge densities [see Eqs.~(\ref{eq:rho-HH})-(\ref{eq:rho-LH})], but
using the $\mbox{\boldmath$\psi$\unboldmath}^h_\alpha$ instead of the rotated
$\mbox{\boldmath$\psi$\unboldmath}^h_{\alpha,R}$. As expected, the lowest subbands are nearly pure HH,
being localized in QW2 (see Fig.~\ref{fig:T-chd-1}), while the $m=7,8$
subbands
are more than 60\% LH. This should not be interpreted as a mixing with
LH states of QW2; rather, it is a manifestation of the fact that this
state is localized in QW1 and, therefore, does not have a well defined
orbital character along [110].

We conclude our analysis of the lowest spin-split doublet by noting
that, as $k_z$ increases, only the lowest level remains a well localized,
quasi-1D state, while its spin companion gradually merges into the QW1
quasi-continuum (at large wavevectors, QW1 becomes
the ground state). For illustration, we show in Fig.~\ref{fig:T-chd-2}
the charge density of the two lowest levels at $k_z =
0.035\,\mbox{\AA}^{-1}$ (close to the wavevector where the QW2 and QW1
dispersions cross). The lowest state (labelled $\uparrow$), which
falls far in energy from the QW bands, is strongly localized at the
intersection of the QWs. The higher state (labelled $\uparrow$) is
peaked at the center of the T but it also extends quite far into QW1.

\subsection{Optical properties}

In Fig.~\ref{fig:T-abs} we show the calculated absorption intensity
for the T-QWR and for light linearly polarized parallel to the wire
axis, $I_\parallel$, and perpendicular to it along the [001]
direction, $I_\perp$. A gaussian broadening of $\sigma_b=\pm 5\,\mbox{meV}$
has been included. We can identify four main
structures, a peak around 1.65 eV, another peak around 1.68 eV (with a
minor shoulder on the low-energy side), a large composite structure
beginning above 1.7 eV, and another large structure at 1.74 eV. Of these,
only the lowest peak involves essentially pure QWR-like localized
states [Fig.~\ref{fig:T-chd-1}(a)]. The second structure involves
transitions from the next electron state, extending in QW1 [see
Fig.~\ref{fig:T-chd-1}(b)], to the first hole states with significant
spatial overlap [see Fig.~\ref{fig:T-chd-1}(d)]. The structure above
1.7 meV involves predominantly QW2 states, with a contributions from
higher QWR resonances which produce the high energy shoulder in
$I_\parallel$. Finally, the large peak at 1.74 eV is due to
many subbands, with a significant contribution from LH states
belonging to both QWs. 
The lowest three structures are polarization dependent with
$I_\parallel > I_\perp$. The anisotropy is maximum for the second
structure, for which $I_\perp$ is very small, consistently with the
expectation for a (001) QW. The higher structure, on the other hand, is
nearly polarization independent, due to the LH contributions.

These results can be compared with the experimental PLE spectra of
Ref.~\onlinecite{Akiyama96a}. There, three main structures occur at
1.64, around 1.67--1.68, and above 1.7 meV. The agreement with the
calculated spectra is again surprisingly good, taking into account
that these neglect excitonic effects. The experimental assignment of
the first structure to a QWR-like state and the successive structures
to QW1-like states, based on the comparison with reference QWs, is
fully consistent with our picture. As concerns intensities, the
agreement is also reasonably good if one considers that the weight of
the QWR peak is sensitive to the relative volume occupied by the QWs,
which enters the calculation through the choice of the supercell. On
the other hand, the reasons of an enhanced intensity of QW1-like
features in the PLE experimental data are discussed in
Ref.~\onlinecite{Akiyama96a}. Finally we compare our results with the
observed values of the anisotropy \cite{Akiyama96a}. From PLE, Akiyama
et al. estimate $I_\perp = 0.39 I_\parallel$ for the QWR peak, and
$I_\perp = 0.14 I_\parallel$ for the next structure assigned to QW1.
The corresponding theoretical values from Fig.~\ref{fig:T-abs} are
approximately $I_\perp = 0.52 I_\parallel$ and $I_\perp = 0.19
I_\parallel$ (the coefficients are slightly larger, 0.57 and 0.17
respectively, if the broadening is reduced to $\sigma_b=\pm
1.5\,\mbox{meV}$). The very different anisotropy of the two structures
is therefore in qualitative agreement with experiments, although 
further investigation would be required to understand
the origin of the difference.

\section*{Summary and conclusions}

We have presented an accurate and efficient approach that allows us to
calculate the electronic and optical properties of quantum wires, taking
into account valence band mixing effects together with realistic profiles
of the confining potentials. We have studied specifically  V- and T-shaped
quantum wires, where the shape of the confinement region differs
considerably from the model geometries assumed in most of the previous
investigations. The two classes of wires differ 
significantly in the structure of their energy spectra: while the crescent 
shape of V-QWRs induces a series of localized quasi-1D levels, only the 
lowest states of T-QWRs are clearly localized in the wire because of 
the subsequent onset of the continua of the parent quantum wells. 
The consequences on the optical spectra have been discussed in detail. 
In particular, we have focused on the optical anisotropy, and 
demonstrated that the analysis of anisotropy spectra can
be used as an effective tool to extract information on valence states,
usually very difficult to obtain otherwise. 

%%% here %%%

Comparison with very recent PLE spectra for both classes of
wires shows good agreement, in spite of our neglecting excitonic
effects. As we discussed in the introduction, this agreement might be due to
the symmetry properties of the Coulomb interaction and, therefore, might be a
rather genereal feature. However, while the approximation of neglecting
excitonic effects is very convenient from the computational point of
view, its accuracy for a given class of materials should be established
{\rm a posteriori} from comparison to experiments, as we have 
positively tested in this paper for V-QWRs and T-QWRs. 

As a final remark, we stress that all our calculations have been
performed by a numerical method which prooved computationally very
convenient. Furthermore, our method lends itself to include
calculations of Coulomb correlation effects on the linear and non
linear optical properties of these wires which are currently
implemented only for non-interacting valence
bands,\cite{Rossi96a} as well as to include external magnetic
fields~\cite{Goldoni95,Bock92b} 
to interpret magneto-luminescence experiments.

%, which are a primary tool of characterization of nanostructures.

\acknowledgments

This work was supported
in part by the EC Commission through the HCM Network ``ULTRAFAST''.

\appendix

\section{Basis set convergence and the choice of $\lowercase{m}^+$, 
$\lowercase{m}^-$}
\label{sec:app-b}

The fictitious masses $m^+$, $m^-$ entering equations
(\ref{eq:fict-f})-(\ref{eq:fict-l}) can be chosen arbitrarily.
In this Appendix we will show that a judicious choice can lead to a
significant improvement in the basis set convergence and lowering of
the computational cost of the calculation. Note that the convergence
with respect to the number of functions $\phi_\nu^+$, $\phi_\nu^-$
included in the basis set, which we investigate below, is a separate
problem from the convergence in the plane-wave expansion of
the  $\phi_\nu^+$, $\phi_\nu^-$ themselves which, for a given structure,
must therefore be checked once and for all {\em before} the diagonalization of
the Luttinger Hamiltonian is started.

In Fig.~\ref{fig:conv} we show the behaviour of the 
lowest hole eigenvalues as a function of the number of 
basis functions $N^+ + N^-$. The calculations have been performed
for the V-QWR A/SL, which is described in detail in section
\ref{sec:V}. The eigenvalues reported in Fig.~\ref{fig:conv} are 
highlighted by black triangles in the full band structure of the same V-QWR
shown in Fig.~\ref{fig:V-bande}: they are the doubly degenerate 
lowest eigenvalue at $k_z=0 $ [panel (a)], and the two spin-split lowest 
eigenvalues at $k_z=0.02\,\mbox{\AA$^{-1}$}$ [panels (b) and (c)]. 

In each panel of Fig.~\ref{fig:conv} we show two sets of 
calculations, both obtained with $m^+=(\gamma_1-2\gamma_2)^{-1}$, but with 
different choices of $m^-$.  The empty dots are obtained 
with $m^-=(\gamma_1+2\gamma_2)^{-1}$. With this choice, $m^+$ and $m^-$ 
are the HH and LH effective masses along the [001] 
crystallographic direction; since this is the direction of strongest 
confinement for these V-QWRs (see next section), these are the 
``physical'' masses in the sense that, for example, 
they would determine the HH and LH 
levels of a QW with comparable confinement length grown in this direction. 
It can be noted, however, that convergence is achieved only within 
$\sim 1\,\mbox{meV}$ with $ N^++N^- $ as high as 280, which corresponds 
to $E^+_{cut}=120\,\mbox{meV}$ and $E^-_{cut}=200\,\mbox{meV}$; these, 
in turn,
should be compared with the low barrier height of this sample, which is 
$85\,\mbox{meV}$. Therefore, one needs to reach energies 
high in the continuum to achieve convergence. The couples of
empty dots at $N^+ + N^-=210$ are obtained with two different choices 
of the pair $(N^+,N^-)$:  $(125,85)$ for the upper points, 
and $(97,113)$ for the lower points. We conclude that, with fixed 
$N^++N^-$, the convergence improves with increasing $N^-$,
suggesting that the $-$ states are responsible for the slow 
convergence. 

In fact, lowering the $m^-$ mass improves the convergence.
In Fig.~\ref{fig:conv}, the black dots correspond to the extremal
choice $m^-=m^+$ which, as we anticipated in Sec.~\ref{sec:metodo}, is our final 
choice, and has  been used in all calculations presented in this paper. 
With respect to the previous case, the convergence is much faster: 
the well-converged values at $N^++N^-=150$ (i.e., $N^+=N^-=75$) are 
obtained with $E^+_{cut}=E^-_{cut}=92.2\,\mbox{meV}$; indeed, we find 
that the 
convergence nearly saturates when $E^+_{cut}=E^-_{cut}\gtrsim85\,\mbox{meV}$, 
i.e., just above the valence band offset for this QWR. 
Note that, in addition to the 
improved convergence,  the choice $m^-=m^+$ implies that
only one of the two equations 
(\ref{eq:fict-f})-(\ref{eq:fict-l}) need to be solved.

As a final remark, we note that the convergence is slower for the 
eigenvalues at $k_z=0.02\,\mbox{\AA$^{-1}$}$ than at $k_z=0$, 
due to the strong HH-LH mixing for large wavevectors. An accurate 
convergence at these wavevectors, as it can be achieved by our 
method, is e.g. necessary to calculate in-wire 
effective mass at the Fermi wavevector. The slow convergence of the in-plane 
effective mass at the Fermi edge is a well-known problem in 
QWs.~\cite{Ando85}

\section{Matrix elements of $\hat{{\bf H}}_L$}
\label{sec:app-a}

Once we have calculated the functions $\phi^+_\nu(x,y)$, $\phi^-_\mu(x,y)$,
we compute the following integrals:
\begin{mathletters}
\begin{eqnarray}
s(\nu,\mu) & = &\int_\Omega \left[\phi^+_\nu(x,y)\right]^* \phi^-_\mu(x,y) \,dx\,dy , 
\\
w_\beta(\nu,\mu) & = &\int_\Omega \left[\phi^+_\nu(x,y)\right]^* \hat{k}_\beta
   \phi^-_\mu(x,y) \,dx\,dy ,
\\
w_{\beta\beta^\prime}(\nu,\mu) & = & 
\int_\Omega \left[\phi^+_\nu(x,y)\right]^* \hat{k}_\beta \hat{k}_{\beta^\prime}
   \phi^-_\mu(x,y) \,dx\,dy ,
\\
v^+_{\beta}(\nu,\nu^\prime) & = &
\int_\Omega \left[\phi^+_\nu(x,y)\right]^* \hat{k}^2_\beta 
   \phi^+_{\nu^\prime}(x,y) \,dx\,dy ,
\\
v^-_{\beta}(\mu,\mu^\prime) & = &
\int_\Omega \left[\phi^-_\mu(x,y)\right]^* \hat{k}^2_\beta 
   \phi^-_{\mu^\prime}(x,y) \,dx\,dy ,
\end{eqnarray}
\end{mathletters}
where $\beta,\beta^\prime\in\{x,y\}$.
In our implementation, the functions $\phi^+_\nu$, $\phi^-_\mu$ are 
expanded in plane waves. Although this is not necessary, it makes very 
easy to compute the above integrals, where the operators $\hat{k}_\beta$ are 
just substituted by scalar numbers $k_\beta$.

With the above definitions, and using the short notations 
$s=s(\nu,\mu)$, $v^+_\beta=v^+_{\beta}(\nu,\nu^\prime)$
$v^-_\beta=v^-_{\beta}(\mu,\mu^\prime)$,
$w_\beta=w_\beta(\nu,\mu)$, and 
$w_{\beta\beta^\prime}=w_{\beta\beta^\prime}(\nu,\mu)$,
the only non-zero matrix elements of $\hat{{\bf H}}_L$ 
in the basis set (\ref{eq:basis-f})-(\ref{eq:basis-l}) are 
\begin{mathletters}
\begin{eqnarray}
\langle\left.+,\nu,\uparrow\right|{\bf H}_L\left|+,\nu^\prime,\uparrow\right.\rangle & = & 
\left[\epsilon^+_\nu+\left(\gamma_1-\frac{\gamma_2-3\gamma_3}{2}\right)k_z^2\right]
\delta_{\nu\nu^\prime} + p^+_x v^+_x + p^+_y v^+_y
\\
\langle\left.+,\nu,\downarrow\right|{\bf H}_L\left|+,\nu^\prime,\downarrow\right.\rangle & = & 
\langle\left.+,\nu,\uparrow\right|{\bf H}_L\left|+,\nu^\prime,\uparrow\right.\rangle
\\
\langle\left.-,\mu,\uparrow\right|{\bf H}_L\left|-,\mu^\prime,\uparrow\right.\rangle & = & 
\left[\epsilon^-_\mu+\left(\gamma_1+\frac{\gamma_2-3\gamma_3}{2}\right)k_z^2\right]
\delta_{\mu\mu^\prime} + p^-_x v^-_x + p^-_y v^-_y
\\
\langle\left.-,\mu,\downarrow\right|{\bf H}_L\left|-,\mu^\prime,\downarrow\right.\rangle & = & 
\langle\left.-,\mu,\uparrow\right|{\bf H}_L\left|-,\mu^\prime,\uparrow\right.\rangle 
\\
\langle\left.+,\nu,\uparrow\right|{\bf H}_L\left|-,\mu,\uparrow\right.\rangle & = & 
-\frac{\sqrt{3}}{2}\left[2\left(\gamma_2 w_{yy}-2i\gamma_3 k_z w_y\right)
-\left(\gamma_2+\gamma_3\right)k_z^2 s -\left(\gamma_2-\gamma_3\right) w_{xx} \right]
\\
\langle\left.+,\nu,\downarrow\right|{\bf H}_L\left|-,\mu,\downarrow\right.\rangle & = & 
-\frac{\sqrt{3}}{2}\left[2\left(\gamma_2 w_{yy}+2i\gamma_3 k_z w_y\right)
-\left(\gamma_2+\gamma_3\right)k_z^2 s -\left(\gamma_2-\gamma_3\right) w_{xx} \right]
\\
\langle\left.+,\nu,\downarrow\right|{\bf H}_L\left|-,\mu,\uparrow\right.\rangle & = & 
2\sqrt{3}\left(\gamma_3 w_{xy}+i\gamma_2 k_z w_x\right)
\\
\langle\left.+,\nu,\uparrow\right|{\bf H}_L\left|-,\mu,\downarrow\right.\rangle & = & 
-2\sqrt{3}\left(\gamma_3 w_{xy}-i\gamma_2 k_z w_x\right)
\end{eqnarray}
\end{mathletters}

\section{Optical transition matrix elements}
\label{sec:app-c}

We define the electron-hole overlap integrals
\begin{eqnarray}
t^+(n,\nu) & = & \int_\Omega \left[\psi^e_n(x,y)\right]^* \phi^+_\nu(x,y)\,dx\,dy, \\
t^-(n,\mu) & = & \int_\Omega \left[\psi^e_n(x,y)\right]^* \phi^-_\mu(x,y)\,dx\,dy. 
\end{eqnarray}

Then, the matrix elements for valence-to-conduction band absorption, with 
light linearly polarized along the [110], [001], and 
[$\overline{1}$10] directions, are the following:

\subsection*{direction [110]} 

\begin{eqnarray}
M_{\alpha\rightarrow n,\uparrow}  
 & = & \sum_{J_m} \langle\left. s,\uparrow\right|p_x\left|3/2,J_m\right.\rangle
  \int\left[\psi^e_n(x,y)\right]^*\psi^h_{J_m}(x,y)\,dx\,dy \nonumber\\
 & = & -\frac{P}{\sqrt{2}}\left[\frac{2}{\sqrt{3}}
      \sum_\mu C^-_\alpha(\mu,\downarrow)t^-(n,\mu)\right]
\end{eqnarray}

\begin{eqnarray}
M_{\alpha\rightarrow n,\downarrow}  
 & = & \sum_{J_m} \langle\left. s,\downarrow\right|p_x\left|3/2,J_m\right.\rangle
  \int\left[\psi^e_n(x,y)\right]^*\psi^h_{J_m}(x,y)\,dx\,dy \nonumber\\
 & = & -\frac{P}{\sqrt{2}}\left[\frac{2}{\sqrt{3}}
      \sum_\mu C^-_\alpha(\mu,\uparrow)t^-(n,\mu)\right]
\end{eqnarray}

\subsection*{direction [001]} 

\begin{eqnarray}
M_{\alpha\rightarrow n,\uparrow}  
 & = & \sum_{J_m} \langle\left. s,\uparrow\right|p_y\left|3/2,J_m\right.\rangle
  \int\left[\psi^e_n(x,y)\right]^*\psi^h_{J_m}(x,y)\,dx\,dy \nonumber\\
 & = & \frac{P}{\sqrt{2}}\left[\sum_\nu C^+_\alpha(\nu,\uparrow)t^+(n,\nu)-
    \frac{1}{\sqrt{3}}\sum_\mu C^-_\alpha(\mu,\uparrow)t^-(n,\mu)\right]
\end{eqnarray}

\begin{eqnarray}
M_{\alpha\rightarrow n,\downarrow}  
 & = & \sum_{J_m} \langle\left. s,\downarrow\right|p_y\left|3/2,J_m\right.\rangle
  \int\left[\psi^e_n(x,y)\right]^*\psi^h_{J_m}(x,y)\,dx\,dy \nonumber\\
 & = & \frac{P}{\sqrt{2}}\left[-\sum_\nu C^+_\alpha(\nu,\downarrow)t^+(n,\nu)+
    \frac{1}{\sqrt{3}}\sum_\mu C^-_\alpha(\mu,\downarrow)t^-(n,\mu)\right]
\end{eqnarray}

\subsection*{direction [$\overline{1}$10]} 

\begin{eqnarray}
M_{\alpha\rightarrow n,\uparrow}  
 & = & \sum_{J_m} \langle\left. s,\uparrow\right|p_z\left|3/2,J_m\right.\rangle
  \int\left[\psi^e_n(x,y)\right]^*\psi^h_{J_m}(x,y)\,dx\,dy \nonumber\\
 & = & \frac{iP}{\sqrt{2}}\left[\sum_\nu C^+_\alpha(\nu,\uparrow)t^+(n,\nu)+
    \frac{1}{\sqrt{3}}\sum_\mu C^-_\alpha(\mu,\uparrow)t^-(n,\mu)\right]
\end{eqnarray}

\begin{eqnarray}
M_{\alpha\rightarrow n,\downarrow}  
 & = & \sum_{J_m} \langle\left. s,\downarrow\right|p_z\left|3/2,J_m\right.\rangle
  \int\left[\psi^e_n(x,y)\right]^*\psi^h_{J_m}(x,y)\,dx\,dy \nonumber\\
 & = & \frac{iP}{\sqrt{2}}\left[\sum_\nu C^+_\alpha(\nu,\downarrow)t^+(n,\nu)+
    \frac{1}{\sqrt{3}}\sum_\mu C^-_\alpha(\mu,\downarrow)t^-(n,\mu)\right]
\end{eqnarray}

\begin{figure}
\caption{Confining potential profiles of V-QWRs and T-QWRs, with
indication of relevant crystallographic directions and reference
frame. For V-QWRs we show two potential profiles,  profile A  (solid
line) and  profile B (dotted line) which are characterized by a
different  value of
the confinement length $L$ (profile A: $L=8.7\,\mbox{nm}$;
profile B: $L=6.83\,\mbox{nm}$) and which will be 
investigated in Sec.~\protect\ref{sec:V}; the supercell periodicity
used in the calculations (see Sec.~\protect\ref{sec:metodo}) is
aproximatly $120\,\mbox{nm}$ along $x$ and $37\,\mbox{nm}$ along $y$.
For the T-QWR, QW1 is truncated at the left hand side at
$x=-50\,\mbox{nm}$; the supercell periodicity is approximately
$55\,\mbox{nm}$ along $x$ and $50\,\mbox{nm}$ along $y$.}
\label{fig:geom}
\end{figure}

\begin{figure}
\caption{Right panel: hole band structure along the free axis 
[$\overline{1}$10] of the V-QWR labelled
A/SL. Black triangles indicate the eigenvalues used in the later
Fig.~\protect\ref{fig:conv} to analize the convergence. Left panel: LH
character of the $k_z=0$ states; the ${\bf J}$ quantization axis is taken
along [001].}
\label{fig:V-bande}
\end{figure}

\begin{figure}
\caption{Total charge density (left panels), and HH- and LH-projected
charge densities (center and right panels) of selected hole subbands,
according to the labels, at $k_z=0$ for the same V-QWR of
Fig.~\protect\ref{fig:V-bande}. Full lines indicate the GaAs/SL
interfaces. For clarity, some charge density maps have been magnified
by a factor of 5 or 10, as indicated by labels. Note that the hole
subbands $m=$ 1, 3, 5, \dots\ at $k_z=0$ are degenerate with the
subbands $m=$ 2, 4, 6, \dots,~\protect\cite{nota-labels} and the total and
projected charge densities are equal for degenerate states.}
\label{fig:V-chd}
\end{figure}

\begin{figure}
\caption{LH character vs. subband energy at $k_z=0$ for samples
with profile A (left panel) and B (right panel), 
and with SL or AlAs barriers (full circles and triangles, respectively).}
\label{fig:sl-vs-bk}
\end{figure}

\begin{figure}
\caption{Optical absorption intensity for linearly polarized light
parallel (thin solid lines) and perpendicular (thin dotted lines) to
the wire axis for the four samples listed in table
\protect\ref{tab:samples}.  The relative anisotropy is also shown
(thick lines), referred to the scale on the right-hand side of each
panel. A gaussian broadening of $\sigma_b=\pm4.5\,\mbox{meV}$ is
included. For each sample, we show a histogram of the strongest
optical transitions at $k_z=0$ and for one spin orientation, for light
polarizion perpendicular the QWR axis. Each bar is proportional in
height to the oscillator strength, and it is labelled with $n/m$,
where $n$ is the index of the conduction state and $m$ the index of
the valence state involved in the transition.~\protect\cite{nota-labels} }
\label{fig:V-abs}
\end{figure}

\begin{figure}
\caption{Energy gap between the hole ground state and the first localized
state with a strong LH character for the V-QWR sample labelled A/SL as
a function of the effective valence band offset $V^h_{\mbox{\tiny
eff}}$. Full dots: full calculation. Empty dots: square well model for
a QW of width $L$ and barrier height  $V^h_{\mbox{\tiny eff}}$; the HH
and LH levels are obtained with $m_{HH}=0.377$) and $m_{LH}=0.090$,
respectively. The experimental value is obtained
as explained in the text.}
\label{fig:e-hh-lh}
\end{figure}

\begin{figure}
\caption{Total charge density of electrons (left panels) and 
holes (right panels)  of the lowest $k_z=0$ subbands 
for the T-QWR.}
\label{fig:T-chd-1}
\end{figure}

\begin{figure}
\caption{Right panel: hole band structure along the free axis [$\overline{1}$10]
of the T-QWR. Solid lines show the dispersion of all states resulting
from the full calculation for our supercell geometry. The full circles
identify the states that, from direct inspection of the wavefunction,
exibit a predominantly localized character at center of the T, and are
therefore assigned to quasi-1D QWR-like states or resonances. As
explained in the text, the remaining states are similar to 2D QW-like
states, with charge density localized mostly in QW1 or QW2. Open
squares and circles show the lowest hole subbands of the parent
isolated QW1 and QW2, respectively. Left panel: LH character (with
${\bf J}$ quantization axis along [110]) of the $k_z=0$ states. }
\label{fig:T-bande}
\end{figure}

\begin{figure}
\caption{Total charge density of the lowest two
hole states of the T-QWR at $k_z=0.035\,\mbox{\AA$^{-1}$}$. The HH-
and LH-projected charge densities of each state are also shown
separately in the center and right-hand panels. Full lines indicate
the GaAs/AlAs interfaces.}
\label{fig:T-chd-2}
\end{figure}

\begin{figure}
\caption{Absorption intensity of the T-QWR for light linearly
polarized parallel (solid line) and perpendicular (dashed line) to the
wire axis along [001], labelled with the main contributions to the
peak intensities. A gaussian broadening of $\pm5\,\mbox{meV}$ is
included.}
\label{fig:T-abs}
\end{figure}

\begin{figure}
\caption{Energy of the lowest eigenvalues for the V-QWR labelled
A/SL (see section \protect\ref{sec:V}) at (a) $k_z = 0$ (spin
degenerate) and (b), (c) at $k_z=0.02\,\mbox{\AA$^{-1}$}$ for the two
spin-split states. Empty dots: eigenvalues calculated with
$m^+=(\gamma_1-2\gamma_2)^{-1}$, $m^-=(\gamma_1+2\gamma_2)^{-1}$. Black dots: eigenvalues
calculated with $m^+=m^-=(\gamma_1-2\gamma_2)^{-1}$. The pairs of empty dots at
$N^++N^-=210$ are calculated with $(N^+,N^-)=(125,85)$ (upper dots)
and $(N^+,N^-)=(97,113)$ (lower dots). }
\label{fig:conv}
\end{figure}

\begin{table}
\caption{Bulk GaAs band parameters used in the calculations}
\label{tab:parameters}
\begin{tabular}{ccccc}
$m_e$ & $\gamma_1$ & $\gamma_2$ & $\gamma_3$ & $E_g$ (eV) \\ \hline
0.067 & 6.85 & 2.1 & 2.9 & 1.519 \\
\end{tabular}
\end{table}

\begin{table}
\caption{V-QWRs sample parameters.}
\label{tab:samples}
\begin{tabular}{ccccc}
	& A/SL	& A/AlAs	& B/SL	& B/AlAs	\\ \hline
L (nm)	& 8.7	& 8.7		& 6.83	& 6.83		\\
Barrier type & (AlAs)$_4$/(GaAs)$_8$ & AlAs & 
               (AlAs)$_4$/(GaAs)$_8$ & AlAs \\
$V^e_{\mbox{\tiny eff}} (eV) $	& 0.150	& 1.036 & 0.150 & 1.036 \\
$V^h_{\mbox{\tiny eff}} (eV) $	& 0.085	& 0.558 & 0.085 & 0.558 \\
\end{tabular}
\end{table}

\begin{table}
\caption{Confinement energies (in meV) of the lowest conduction and
valence states at $k_z=0$ for the V-QWRs.  Note that at $k_z=0$ spin
degeneracy holds. Therefore, subband $m=1$ is degenerate to $m=2$, etc.
\protect\cite{nota-labels}}
\label{tab:V-edges}
\begin{tabular}{ccccc}
\multispan{5}{\hfil Electrons \hfil} \\ 
 n	& A/SL	& A/AlAs	& B/SL	& B/AlAs	\\ \hline
 1	& 43.3	& 68.3	& 55.5	& 98.8	\\
 2	& 57.3	& 91.9	& 63.2	& 117.0	\\
 3 	& 65.3	& 112.5	& 67.0	& 124.0	\\
 4	& 69.8	& 126.1	& 72.4	& 132.2 \\ \hline
\multispan{5}{\hfil Holes \hfil} \\ 
 m	& A/SL	& A/AlAs	& B/SL	& B/AlAs	\\ \hline
 1,2    & 10.7  & 14.4  & 13.1  & 18.7 \\
 3,4    & 13.1  & 18.1  & 13.8  & 19.5 \\
 5,6    & 14.4  & 20.0  & 15.6  & 21.7 \\
 7,8    & 15.8  & 21.8  & 17.1  & 23.7
\end{tabular}
\end{table}

% \begin{table}
% \caption{}
% \label{}
% \begin{tabular}{}
% \end{tabular}
% \end{table}


\begin{references} 
\bibitem{review} For a review see: R. Cingolani and R. Rinaldi, 
Rivista Nuovo Cimento {\bf 16}, 1 (1993).
\bibitem{Kapon89} E. Kapon, D.M. Hwang, 
and R. Bhat, Phys. Rev. Lett. {\bf 63}, 430 (1989).
\bibitem{Gailhanou93} M. Gailhanou {\em et al.},
Appl. Phys. Lett. {\bf 62}, 1623 (1993).
\bibitem{Tiwari94} S. Tiwari {\em et al.}, 
Appl. Phys. Lett. {\bf 64}, 3536 (1994).
\bibitem{Rinaldi94a} R. Rinaldi, {\em et al.}, Phys.~Rev.~B~{\bf 50},
11795 (1994).
\bibitem{Grundmann95} M. Grundmann, {\em et al.}, J. Nonlin. Optical Phys. 
and Materials {\bf 4}, 99 (1995).

\bibitem{Pfeiffer90} L.N. Pfeiffer {\em et al.}, 
Appl. Phys. Lett. {\bf 56}, 1697 (1990).
\bibitem{Gershoni90} D. Gershoni {\em et al.}, 
Phys. Rev. Lett. {\bf 65}, 1631 (1990).
\bibitem{Stormer91} H.L. St\"ormer {\em et al.}, 
Appl. Phys. Lett. {\bf 58}, 726 (1991).
\bibitem{Goni92} A.R. Go\~ni {\em et al.}, 
Appl. Phys. Lett. {\bf 61}, 1956 (1992).
\bibitem{Wegscheider95} W. Wegscheider {\em et al.}, 
J. Cryst. Growth {\bf 150}, 285 (1995).

\bibitem{Akiyama96a} H. Akiyama, T. Someya, and H. Sakaki, 
Phys.~Rev.~B {\bf 53}, R4229 (1996).
\bibitem{Akiyama96b} H. Akiyama, T. Someya, and H. Sakaki,
Phys.~Rev.~B {\bf 53}, R10520 (1996).
\bibitem{Someya96} T. Someya, H. Akiyama, and H. Sakaki,
Phys.~Rev.~Lett {\bf 76}, 2965 (1996).

\bibitem{Sercel90} P.C. Sercel and K.J. Vahala, App. Phys. Lett. {\bf
57}, 545 (1990).
\bibitem{Citrin89} D.S. Citrin and Y.-C. Chang, Phys.~Rev.~B {\bf 40}, 
5507 (1989).
\bibitem{Bock92a} U. Bockelmann and G. Bastard, Europhys. Lett.
{\bf 15}, 215 (1991); Phys.~Rev.~B {\bf 45}, 1688 (1992).
\bibitem{Citrin92} D.S. Citrin and Y.-C. Chang, Phys.~Rev.~B {\bf 43}, 
11703 (1992).
\bibitem{Ando93} H. Ando, S. Nojima, and H. Kanbe, J. Appl. Phys.{\bf 74},
6383 (1993).
\bibitem{Goldoni95} G. Goldoni and A. Fasolino, Phys.~Rev.~B {\bf 52}, 
14118 (1995); Physica B {\bf 211}, 444 (1995).
\bibitem{Yamaguchi95} A.A. Yamaguchi and A. Usui,
J. Appl. Phys. {\bf 78}, 1361 (1995).
\bibitem{Rossi96a} F. Rossi and E. Molinari, Phys.~Rev.~Lett.~{\bf76}, 
3642 (1996); Phys.~Rev.~B {\bf 53}, 16462 (1996); F. Rossi and E.
Molinari, in ``The Physics of Semiconductors'', edited by M. Scheffler and R.
Zimmermann (World Scientific, Singapore 1996), p. 1161.
\bibitem{our-apl} G. Goldoni, F. Rossi, E. Molinari, A. Fasolino, R. 
Rinaldi, and R. Cingolani, Appl.~Phys.~Lett.~in press.
\bibitem{nota-labels} Note that the
electron subband index $n$ does not include spin, and two spin orientations
correspond to each $n$, while for holes the index $m$ includes the 
spin character. 
\bibitem{Luttinger56} J. M. Luttinger, Phys.~Rev.~{\bf 102}, 1030 (1956).
\bibitem{Xia91} J.-B. Xia, Phys.~Rev.~B {\bf 43}, 9856 (1991).

\bibitem{Rinaldi94b} R. Rinaldi, {\em et al.}, Phys.~Rev.~Lett.~{\bf 73},
2899 (1994).

\bibitem{nota-spin} For holes, the word ``spin'' is used in a 
generalized way, and does not correspond to a well defined spin 
orientation. %CHECK
See D.A. Broido and L.J. Sham, Phys.~Rev.~B~{\bf 31}, 
888 (1985); L.C. Andreani, A. Pasquarello, and F. Bassani, Phys.~Rev.~B~{\bf
36}, 5887 (1987); G. Goldoni and A. Fasolino, Phys.~Rev.~Lett.~{\bf 69},
2567 (1992); {\bf 69}, 2567 (1992).
\bibitem{Los95} The QW subbands are calculated by a finite element solution
of the Luttinger Hamiltonian in the relevant crystallographic
directions. See J. Los, 
A. Fasolino, and A. Catellani, Phys. Rev. B {\bf 53}, 4630
(1996). Note that the QWR ground state is actually sligtly above the
QW2 continuum, contrary to the expectation for a localized level.
Actually, this is a fictitious result due to the different accuracy
between the plane wave expasion of QWR states and the real-space
representation used to calculate the QW states. In reality, the QWR
ground state, being localized, lies slightly below the QW continuum.
However, for this structure the binding energy is very small.
\bibitem{Bastard} G. Bastard, {\em Wave mechanics applied to
semiconductor heterostructures} (Les \'editions des physique, Les
Ulis, Paris, France 1988).
\bibitem{Bock92b} U. Bockelmann and G. Bastard, Phys.
~Rev.~B {\bf 45}, 1700 (1992).
\bibitem{Ando85} T. Ando, J. Phys. Soc. Japan {\bf 54}, 1528 (1985).

\end{references}
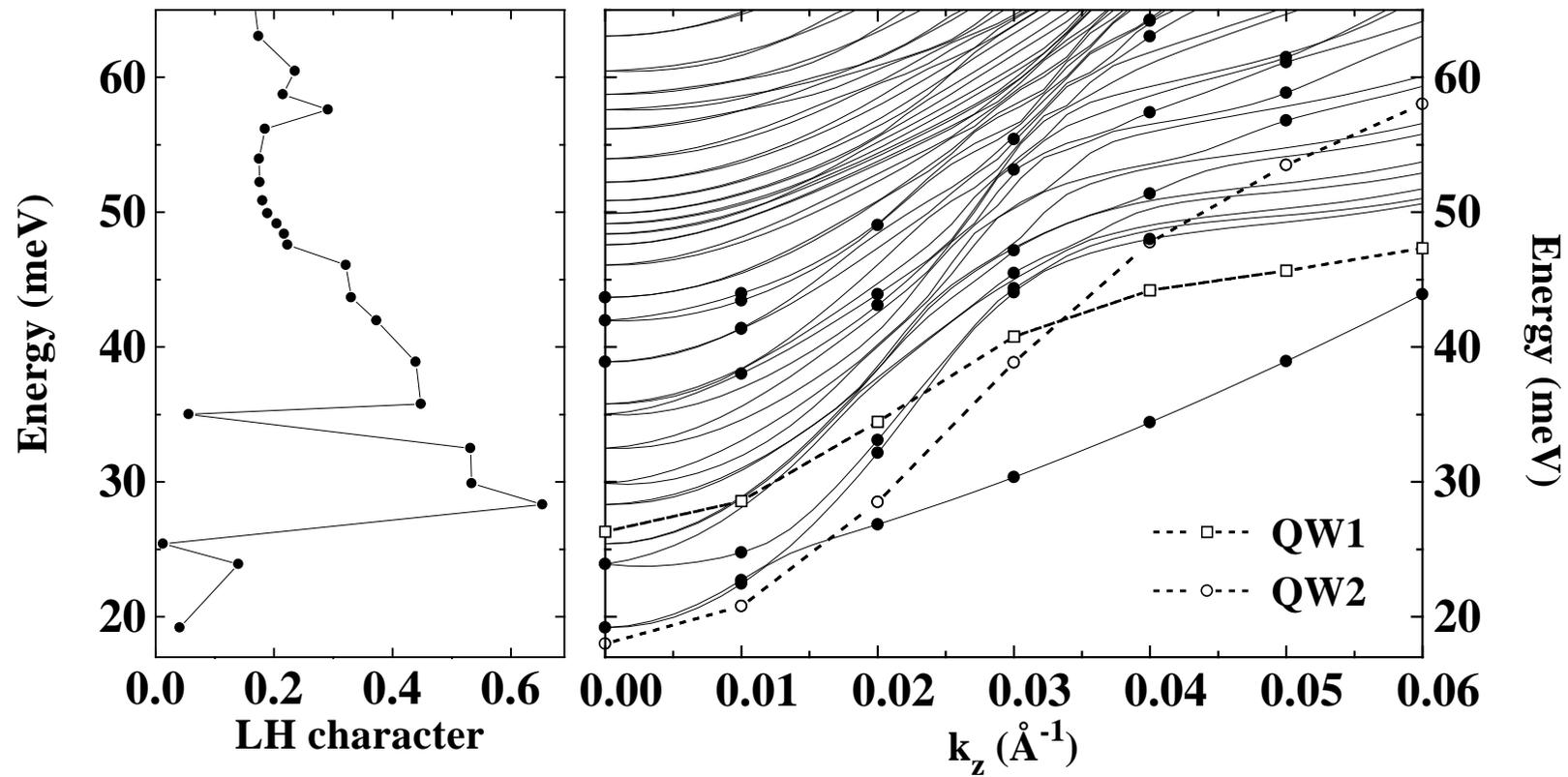
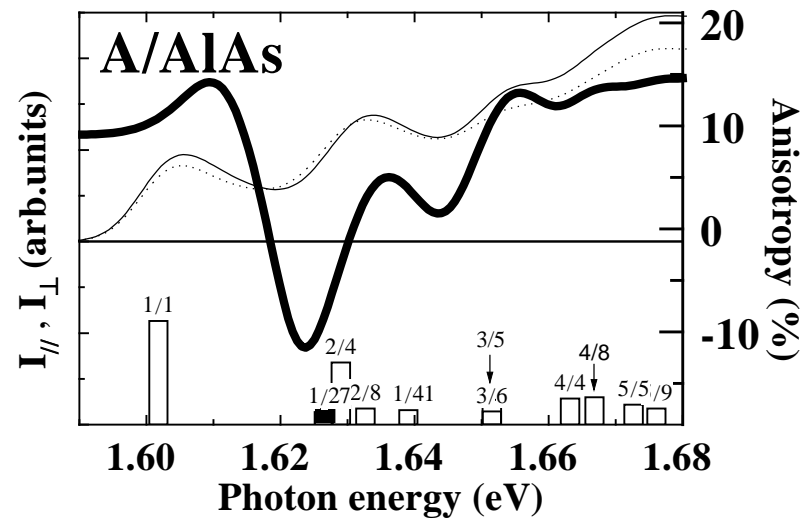
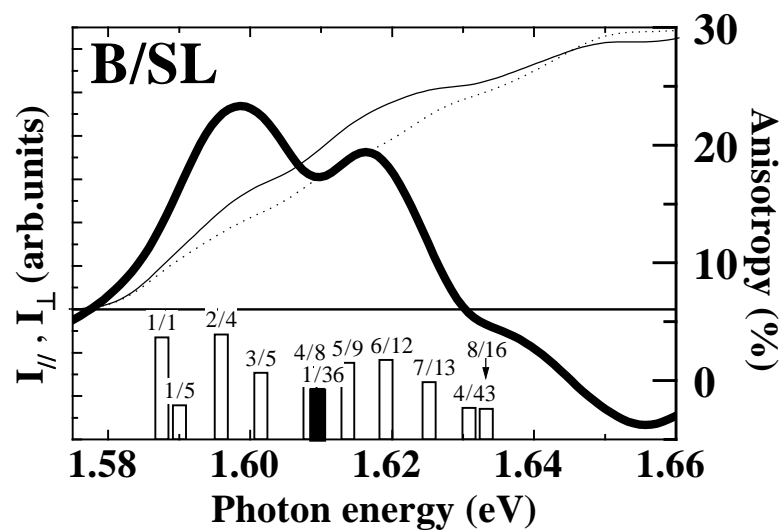
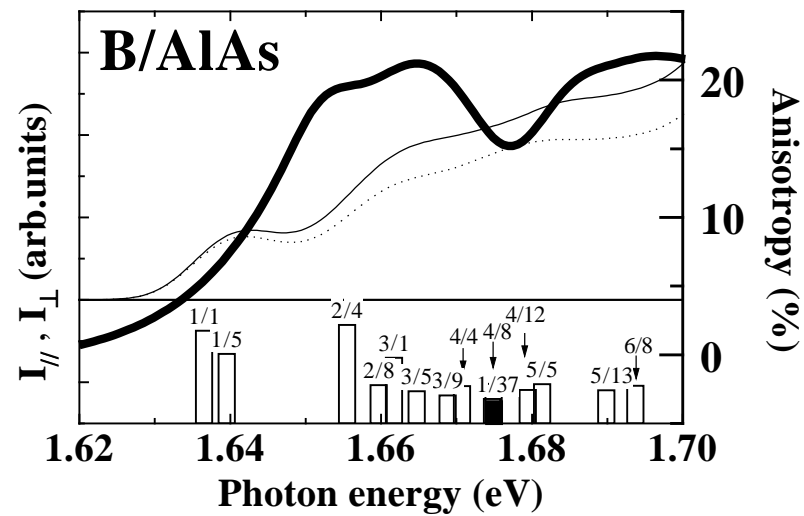
\end{document}